\documentclass[twocolumn,showpacs,amssymb,amsmath,aps,floatfix,prl,nofootinbib,superscriptaddress]{revtex4-1}

\usepackage{hyperref}
\usepackage[all]{hypcap}
\usepackage{array}
\usepackage{graphicx}
\usepackage{amsmath,amsthm,amssymb}

\usepackage{color}

\begin{document}

\title{Bayesian Extraction of Jet Energy Loss Distributions in Heavy-Ion Collisions}

\author{Yayun He}

\affiliation{Key Laboratory of Quark \& Lepton Physics (MOE) and Institute of Particle Physics, Central China Normal University, Wuhan 430079, China}
\affiliation{Nuclear Science Division, Lawrence Berkeley National Laboratory, Berkeley, California 94720, USA}

\author{Long-Gang Pang}
\email{lgpang@lbl.gov}
\affiliation{Nuclear Science Division, Lawrence Berkeley National Laboratory, Berkeley, California 94720, USA}
\affiliation{Physics Department, University of California, Berkeley, California 94720, USA}

\author{Xin-Nian Wang}
\email{xnwang@lbl.gov}
\affiliation{Key Laboratory of Quark \& Lepton Physics (MOE) and Institute of Particle Physics, Central China Normal University, Wuhan 430079, China}
\affiliation{Nuclear Science Division, Lawrence Berkeley National Laboratory, Berkeley, California 94720, USA}
\affiliation{Physics Department, University of California, Berkeley, California 94720, USA}

\begin{abstract}
Based on the factorization  in perturbative QCD,  a jet cross sections in heavy-ion collisions can be expressed as a convolution of the jet cross section in $p+p$ collisions and a jet energy loss distribution. Using this simple expression and the Markov Chain Monte Carlo method, we carry out Bayesian analyses of experimental data on jet spectra to extract energy loss distributions for both single inclusive and $\gamma$-triggered jets in $Pb+Pb$ collisions with different centralities at two colliding energies at the Large Hadron Collider.  The average jet energy loss has a dependence on the initial jet energy that is slightly stronger than a logarithmic form and decreases from central to peripheral collisions. The extracted jet energy loss distributions with a scaling behavior in $x=\Delta p_T /\langle \Delta p_T\rangle$ have a large width. These are consistent with the linear Boltzmann transport model simulations, in which the observed jet quenching is caused on the average by only a few out-of-cone scatterings.
\end{abstract}



\maketitle

\noindent{\it Introduction}.--Suppression of jets and large transverse momentum hadrons known as jet quenching~\cite{Gyulassy:1990ye,Wang:1991xy} in high-energy heavy-ion collisions is caused by the interaction between jet-shower and medium partons and can be used to probe properties of the quark-gluon plasma (QGP). One such fundamental property of QGP is the jet transport coefficient \cite{Baier:1996sk} which characterizes the average transverse momentum broadening squared per unit length of a propagating parton and is directly related to the gluon distribution density of the QGP medium~\cite{CasalderreySolana:2007sw}. Its value at the energy scale of thermal momentum is also related to the shear viscosity of the QGP \cite{Majumder:2007zh}.  Among many efforts to extract the jet transport coefficient from experimental data on suppression of single inclusive hadron spectra~\cite{Adare:2008cg,Bass:2008rv,Armesto:2009zi,Chen:2010te},  the systematic study by the JET Collaboration \cite{Burke:2013yra} has narrowed the uncertainties to within 40\%. Such a systematic approach has still yet to be applied to other experimental measurements of jet quenching, such as suppression of single inclusive and $\gamma$-triggered jets.

The study of the medium modification of fully reconstructed jets in high-energy heavy-ion collisions~\cite{Vitev:2008rz,Vitev:2009rd} can provide additional constraints on theoretical approaches to parton energy loss and the jet transport coefficient. Though a fully constructed jet contains partons both from the medium-modified jet shower and medium recoil~\cite{Wang:2013cia,Tachibana:2015qxa,Casalderrey-Solana:2016jvj,Wang:2016fds,Tachibana:2017syd,KunnawalkamElayavalli:2017hxo,Milhano:2017nzm,Chen:2017zte,Luo:2018pto}, one can still define jet energy loss as the difference between the final jet energies within the jet cone in vacuum and medium originating from the same initial hard parton. While the average jet energy loss is related to both jet and bulk transport coefficients, the jet energy loss distribution should contain additional information about jet-medium interaction. It is important to extract both from experimental data.


In this Letter, we first show that starting from the factorized form of jet cross section, the jet production cross section in heavy-ion collisions can be expressed as the convolution of cross section in proton-proton collisions and a flavor-averaged jet energy loss distribution. Based on this simple expression, we use the Markov Chain Monte Carlo (MCMC) method \cite{Andrieu:2003} to carry out the first Bayesian analyses of experimental data on the medium modification of both single inclusive and $\gamma$-triggered jet spectra and extract jet energy loss distributions in heavy-ion collisions at two colliding energies at the Large Hadron Collider (LHC) with different centralities. Previous efforts have been carried out to extract the averaged jet and parton energy loss from suppression of single inclusive jet~\cite{Spousta:2015fca} and hadron spectra \cite{Adcox:2004mh,Arleo:2017ntr}  based on either a simple average energy loss or one particular model for parton energy loss distribution. Our study in this Letter uses Bayesian analyses with uniform prior distributions of parameters to extract the jet energy loss distribution assuming a $p_T$-dependent average jet energy loss and a scaling behavior of the jet energy loss distribution.

\noindent{\it Jet production cross section}.--Within the factorized parton model in perturbative QCD (pQCD), the double differential cross section for single inclusive jet production in $p+p$ collisions can be factorized \cite{Kang:2016mcy,Kang:2017frl},
\begin{equation}
\frac{d\sigma_{pp}^{\rm jet}}{dp_Td\eta}=\sum_{a,b,c}\int f_{a/{\rm p}}\otimes f_{b/{\rm p}}\otimes H_{ab}^{c}\otimes J_c(p_T, R|p_{Tc}),
\end{equation}
as the convolution of parton distribution functions $f_{a/p}$, hard functions $H_{ab}^c$ for parton scattering subprocesses $a+b\rightarrow c+X$ and semi-inclusive jet functions $J_c(p_T, R|p_{Tc})$ which describe the formation of a jet with transverse energy $p_T$ and jet-cone size $R$ from a parent parton $c$ with initial transverse momentum $p_{Tc}$.  Similarly, the cross section in $A+A$ collisions is given by
\begin{eqnarray}
\frac{d \sigma^{\rm jet}_{AA}}{dp_{T}d\eta} & = &\sum_{a,b,c}  \int d^2{\bf r} d^2{\bf b}  t_A(r) t_A(|{\bf b}-{\bf r}|) \frac{d\phi_c}{2\pi}   \nonumber\\
&& \hspace{-0.4in} \times f_{a/{A}} \otimes f_{b/{A}} \otimes H_{ab}^c \otimes  \widetilde{J}_{c}(p_T,R,{\bf r},{\bf b},\phi_c|p_{Tc}),
\label{eq:cs.aa}
\end{eqnarray}
where $t_{A}(r)$ is the nuclear thickness function normalized to $A$, $f_{a/A}$ is the parton distribution function per nucleon inside the nucleus~\cite{Eskola:2009uj}, $\bf r$ is the transverse coordinate of the binary nucleon-nucleon collision that produces the initial hard parton $c$ with the azimuthal angle $\phi_c$, $\bf b$ is the impact parameter of the nucleus-nucleus collision and $\widetilde{J}_{c}$ is the medium-modified semi-inclusive jet function. The range of integration over the impact parameter $\bf b$ is determined by the centrality of the nucleus-nucleus collisions according to experimental measurements.

Interaction between jet shower and medium partons during the jet transport through the QGP medium will lead to transverse diffusion of jet shower partons to the outside of the jet cone. Radiated gluons and medium response, on the other hand, can also fall into the jet cone. These will lead to an effective jet transverse energy loss which we define as the difference between the jet transverse energies in vacuum and medium originating from the same initial parton $c$. Note this jet energy loss is very different from the energy loss of an individual parton. For a given ${\bf r}$, ${\bf b}$, and $\phi_c$, the medium-modified jet function can be given as the convolution, 
\begin{eqnarray}
 \widetilde{J}_{c}(p_T,R,{\bf r},{\bf b},\phi_c|p_{Tc}) &=&\int d\Delta p_T  J_c(p_T+\Delta p_T, R|p_{Tc}) \nonumber \\
&&\hspace{-0.5in} \times w_c(\Delta p_T, p_{T}+\Delta p_T, R, {\bf r},{\bf b},\phi_c),
\end{eqnarray}
of vacuum jet function with transverse energy $p_T+\Delta p_T$ and a $p_T$-dependent jet energy loss distribution $w_c$.

Averaging over the parton production point and propagation direction, the cross section for single inclusive jet production in $A+A$ collision in Eq.~(\ref{eq:cs.aa}) can be written as
\begin{eqnarray}
\frac{d \sigma^{\rm jet}_{AA}}{dp_{T}d\eta} & = &N_{\rm bin}(b)\sum_{a,b,c}  \int  d\Delta p_T W^{c}_{AA}(\Delta p_T, p_{T}+\Delta p_T,R) \nonumber\\
&& \hspace{-0.2in} \times f_{a/{A}} \otimes f_{b/{A}} \otimes H_{ab}^c \otimes  J_{c}(p_T+\Delta p_T,R|p_{Tc}),
\label{eq:cs.aa2}
\end{eqnarray}
where $N_{\rm bin}(b)=\int d^2{\bf r} d^2{\bf b}  t_A(r) t_A(|{\bf b}-{\bf r}|)$ is the number of binary collisions and  the energy loss distribution for a given centrality class of $A+A$ collisions is defined as
\begin{eqnarray}
W^{c}_{AA}(\Delta p_T, p_{T}, R)&=&\int d^2{\bf r} d^2{\bf b}  t_A(r) t_A(|{\bf b}-{\bf r}|) \frac{d\phi_c}{2\pi} \nonumber \\
&\times& \frac{w_c(\Delta p_T, p_{T}, R, {\bf r},{\bf b},\phi_c)}{N_{\rm bin}(b)}.
\end{eqnarray}
For jet production at a very high transverse energy, one can neglect nuclear modification of parton distribution functions $f_{a/{A}}\approx f_{a/p}$. The single inclusive jet cross section in $A+A$ collisions can be expressed as the convolution of jet cross section in $p+p$ collisions and a flavor-averaged (quarks and gluon) jet energy loss distribution
$W_{AA}$. The modification factor for single inclusive jet production in $A+A$ collisions can be written as,
\begin{eqnarray}
 R_{AA}(p_{T}) &\approx& \frac{1}{d\sigma^{\rm jet}_{pp}(p_T)} \int d\Delta p_T d\sigma^{\rm jet}_{pp}(p_{T} + \Delta p_T) \nonumber\\
&\times& W_{AA}(\Delta p_T, p_T+\Delta p_T, R).
 \label{eq:raa}
\end{eqnarray}
This expression for single inclusive jet cross section should also be valid for $\gamma$-triggered jet spectra and has been postulated~\cite{Spousta:2015fca,Mehtar-Tani:2017web} before. Similar approximate expression for single inclusive hadron spectra has been used in Refs.~\cite{Arleo:2017ntr,Baier:2001yt} assuming a constant average momentum fraction of hadrons $z_h=p_{Th}/p_T$ in the energy loss distribution. This approximation is, however, no longer valid for hadron- and $\gamma/Z^0$-triggered hadron spectra and jet fragmentation functions where hadrons from radiated gluon and medium response have to be included.

\noindent{\it Bayesian analyses with {\rm MCMC}}.--The focus of the rest of this Letter is to use Bayesian analyses of experimental data on both single inclusive and $\gamma$-triggered jet spectra in $p+p$ and $A+A$ collisions to extract the jet energy loss distribution $W_{AA}$ using the convolution expression in Eq.~(\ref{eq:raa}).  For this purpose we have to assume a general functional form for $W_{AA}$.  The average jet energy loss $\langle \Delta p_T\rangle \equiv \int d\Delta p_T \Delta p_T W_{AA}(\Delta p_T, p_T, R)$ should be a function of the vacuum or initial jet energy $p_T$ for a given jet-cone size $R$. The fluctuation of the energy loss for a given centrality class of $A+A$ collisions is mainly determined by the variation in the number of jet-medium scatterings according to the distribution of the initial jet production position and azimuthal angle of the propagation. 
Motivated by results from linear Boltzmann transport (LBT) simulations~\cite{He:2018xjv}, we assume the jet energy loss distribution $W_{AA}(\Delta p_T, p_T,R)\approx W_{AA}(x, R)$ is approximately a function of the scaled jet energy loss $x=\Delta p_T/\langle \Delta p_T\rangle$. The dependence of $W_{AA}$ on vacuum jet energy $p_T$ is only implicit through the average jet energy loss $\langle \Delta p_T\rangle (p_T)$.  We will take the scaled energy loss distribution as a normalized $\Gamma$ function,
\begin{equation}
W_{AA}(x)=\frac{\alpha^\alpha x^{\alpha-1}e^{-\alpha x}}{\Gamma(\alpha)},
\label{eq:waa}
\end{equation}
where $\alpha$ is approximately independent of the jet energy. When $\alpha$ is an integer, the $\Gamma$ function is in fact a convolution of $\alpha$ number of exponential distributions. One can empirically interpret the above energy loss distribution as a consequence of $\alpha$ number of jet-medium scatterings that transport partons out of the jet cone with an average jet energy loss per out-of-cone scattering $\langle \Delta p_T\rangle/\alpha$.

The jet energy dependence of the average jet energy loss is assumed to have the following functional form,
\begin{equation}
\langle \Delta p_T\rangle (p_T)=\beta p_T^\gamma \log(p_T),
\label{eq:eloss}
\end{equation}
since the energy loss should be zero at $p_T=0$. The log term is motivated by theoretical calculations of parton energy loss~\cite{He:2015pra}. The problem is then reduced to estimation of three parameters $[\alpha,\beta,\gamma]$ through Bayesian analyses of experimental data with MCMC method.
 
Bayesian analyses have been employed to extract bulk and heavy quark transport coefficients~\cite{Pratt:2015zsa,Bernhard:2016tnd,Xu:2017obm} in heavy-ion collisions from comparisons between experimental data and model simulations. We use the same statistical analysis to extract probability distributions of the parameters in the jet energy loss distribution from experiment data. The process can be summarized as
\begin{equation}
P(\theta |{\rm data}) = \frac{P(\theta) P({\rm data} | \theta)}{P({\rm data})},
\label{eq:bayesian}
\end{equation}
 where $P(\theta | {\rm data})$ is the posterior distribution of parameters $\theta=[\alpha,\beta,\gamma]$ given the experimental data,
 $P(\theta)$ is the prior distribution of $\theta$, $P({\rm data} | \theta)$ is the Gaussian likelihood between experimental data
 and the output for any given set of parameters $\theta$ and $P({\rm data}) = \int d\theta P(\theta) P({\rm data} | \theta)$ is 
 the evidence. Uncorrelated uncertainties in experimental data are used in the evaluation of the Gaussian likelihood.  The MCMC \cite{Andrieu:2003} method is a strategy for generating $\theta$, whose distribution mimics an unnormalized probability distribution $\propto P(\theta) P({\rm data} | \theta)$, with importance sampling. 
 
 \begin{figure}
\centerline{\includegraphics[width=8.5cm]{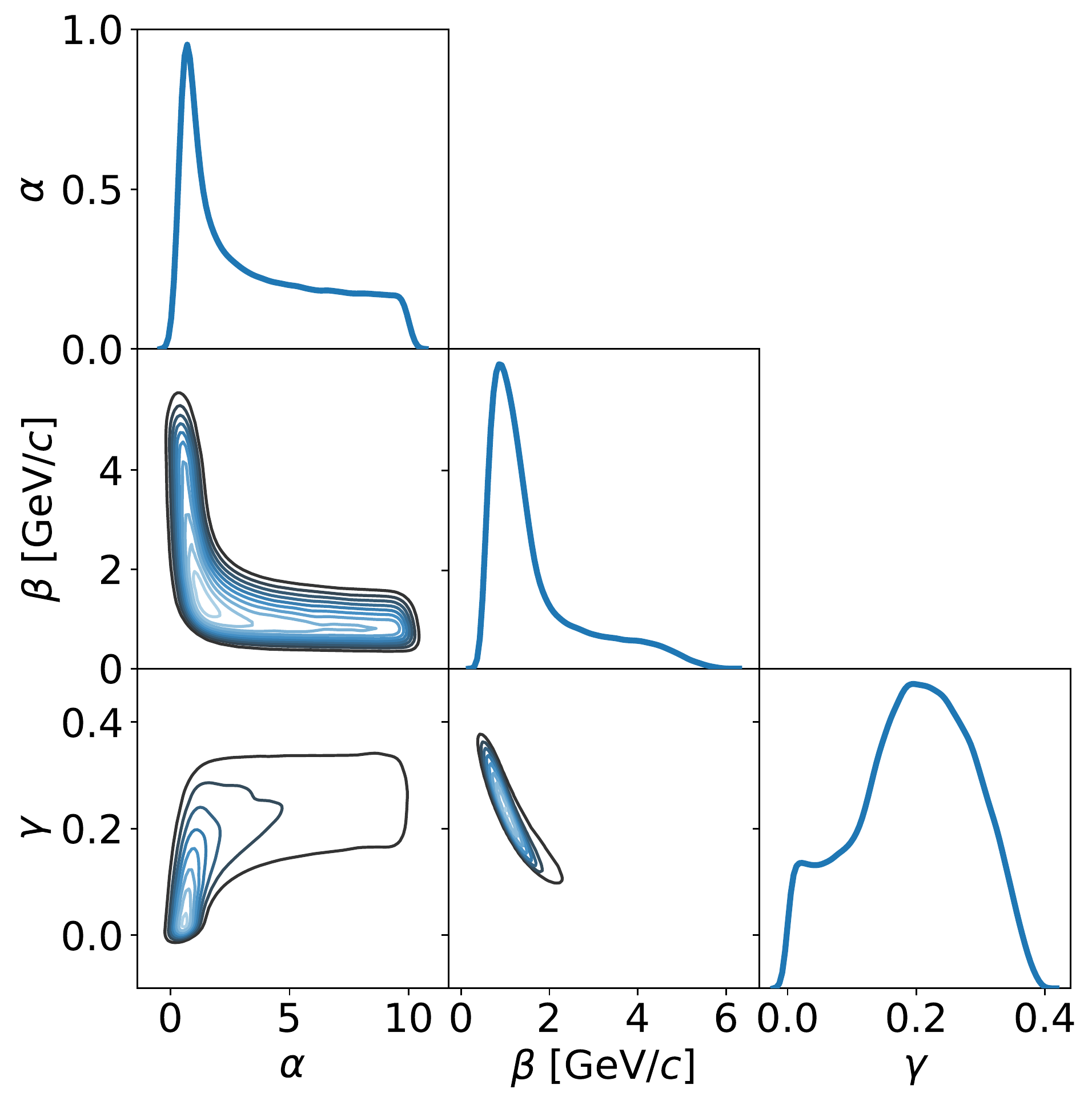}}
 \caption{Density distributions and correlations of the parameters $\theta =[\alpha,\beta,\gamma]$ in the jet energy loss distribution from Bayesian fit to experimental data on single inclusive jet suppression in 0-10\% central $Pb+Pb$ collisions at $\sqrt{s}=2.76$ TeV.}
 \label{fig:parameter}
\end{figure}
 
 In this Letter, PyMC library \cite{huard,github} is employed to carry out the MCMC estimation of the parameters for the jet energy loss distribution with Metropolis-Hastings random walk in the parameter space. The maximum {\it a posterior} (MAP) method is used first in PyMC to get a fast estimation of the parameters $\theta$. The estimated values are fed as the initial guess of these parameters in MCMC to sample $16\times 10^6$ sets in the parameter space. The first $8\times 10^6$ samples are treated as burn-in samples which are not used in the final statistics.  
 
\noindent {\it Jet energy loss distributions}.--The experimental data used for the Bayesian analyses in this study are from ATLAS for single inclusive jet spectra~\cite{Aad:2014bxa,Aaboud:2018twu} and CMS for $\gamma$-triggered jet spectra~\cite{Chatrchyan:2012gt,CMS:2013oua,Sirunyan:2017qhf} in $p+p$ and $Pb+Pb$ collisions at $\sqrt{s}=2.76$ and 5.02 TeV. The jet spectra in $p+p$ collisions beyond the experimental $p_T$ range are provided by PYTHIA simulations \cite{Sjostrand:2007gs}. A uniform prior distribution $P(\theta)$ in the region $[\alpha,\beta,\gamma]\in [(0,10),(0,10),(0,1)]$ is used for the Bayesian analyses. We have also tried normal prior distributions with model-motivated means and large variances. The results remain the same as that with uniform prior distributions. We check the convergence of the analyses by examining the density distributions and pair correlations of the parameters, as shown in Fig.~\ref{fig:parameter}, for example for the fit to $R_{AA}$ for single inclusive jet spectra in 0-10\% $Pb+Pb$ collisions at $\sqrt{s}=2.76$ TeV. Clearly, the three parameters in the jet energy loss distributions from the Bayesian fits are strongly correlated.

\begin{figure}
\centerline{\includegraphics[width=8.5cm]{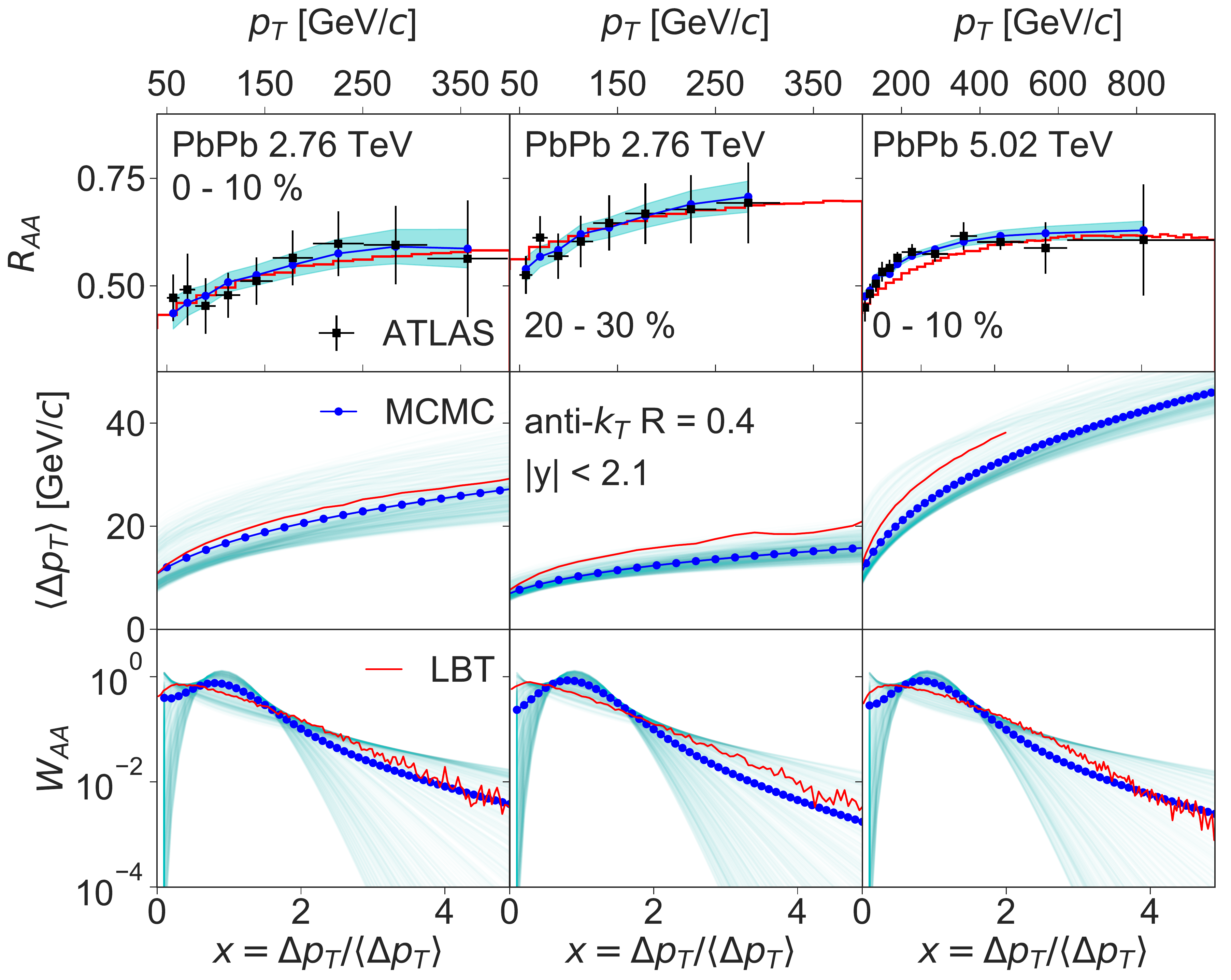}}
 \caption{ (Top) Bayesian fits to $R_{AA}$ for single inclusive jets \cite{Aad:2014bxa,Aaboud:2018twu}, (middle) the extracted average jet energy loss  $\langle \Delta p_T\rangle$ as a function of the initial jet energy and (bottom) energy loss distributions $W_{AA}(x=\Delta p_T/\langle \Delta p_T\rangle)$  in $Pb+Pb$ collisions at two LHC energies with different centralities. Blue lines with solid circles are mean averages from MCMC Bayesian fits and light blue lines are results with one sigma deviation from the average fits of $R_{AA}$.  Red lines are from LBT simulations.}
 \label{fig:singlejet}
\end{figure}

\begin{figure}
\centerline{\includegraphics[width=8.5cm]{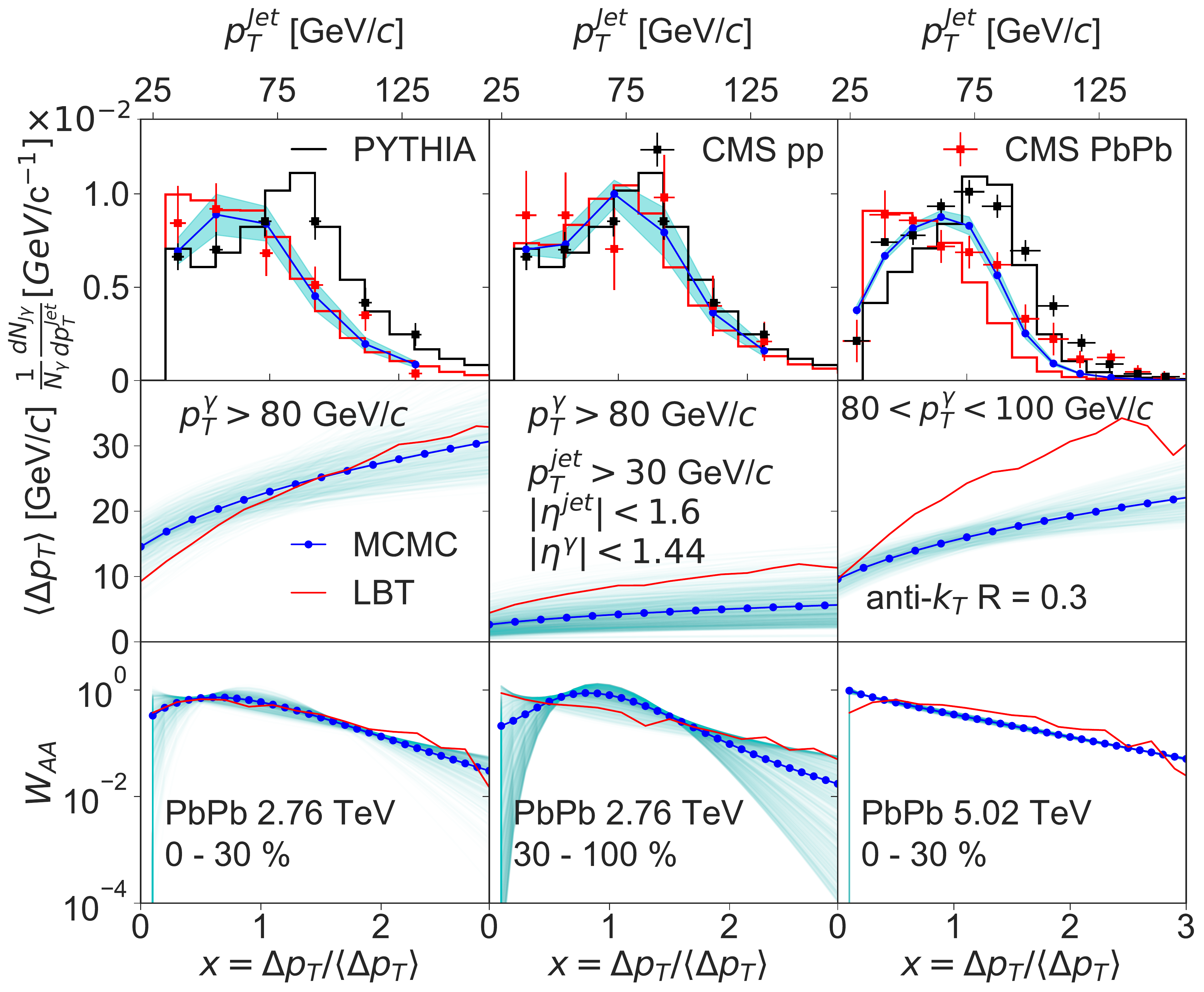}}
 \caption{The same as Fig.~\ref{fig:singlejet} except for fits to $\gamma$-triggered jet spectra~\cite{Chatrchyan:2012gt,CMS:2013oua,Sirunyan:2017qhf} in $Pb+Pb$ collisions at (left and middle) $\sqrt{s}=2.76$  and (right) 5.02 TeV.}
 \label{fig:gammajet}
\end{figure}

Shown in Fig.~\ref{fig:singlejet} are the final fits (top panel) to the ATLAS data \cite{Aad:2014bxa,Aaboud:2018twu} on $R_{\rm AA}$ of single inclusive jets,  the extracted average jet energy loss  $\langle \Delta p_T\rangle$ as a function of the vacuum jet energy $p_T$ (middle panel) and energy loss distributions $W_{AA}(x=\Delta p_T/\langle \Delta p_T\rangle)$ (bottom panel) in 0-10\% (left column), 20-30\% (middle column) $Pb+Pb$ at $\sqrt{s}=2.76$ TeV and 0-10\% $Pb+Pb$ collisions at $\sqrt{s}=5.02$ TeV (right column). The blue lines with solid circles are mean averages from the MCMC Bayesian fits and the light blue lines are results with one sigma deviation from the average fits of $R_{\rm AA}$.  Similarly in Fig.~\ref{fig:gammajet}, we show the fits (top panel) to the CMS data \cite{Chatrchyan:2012gt,CMS:2013oua,Sirunyan:2017qhf} on medium-modified $\gamma$-jet spectra $(1/N_\gamma)dN_{j\gamma}/dp_T$, the extracted average jet energy loss as a function of the vacuum jet energy (middle panel) and energy loss distributions (bottom panel) in 0-30\% (left column), 30-100\% (middle column) Pb+Pb at $\sqrt{s}=2.76$ TeV and 0-30\% Pb+Pb collisions at $\sqrt{s}=5.02$ TeV. We can see that the Bayesian fits of single inclusive and $\gamma$-triggered jet spectra based on the convolution expression in Eq.~(\ref{eq:raa}) can describe the experimental data very well. The extracted jet energy loss distributions as parametrized in Eqs.~(\ref{eq:waa}) and (\ref{eq:eloss}) are quite broad and average jet energy loss increases with the vacuum jet energy a little faster than a logarithmic dependence. The parameters (mean and one-$\sigma$ variance in fits to $R_{AA}$ and $\gamma$-triggered jet spectra) for jet energy loss distributions from the Bayesian fits in Figs.~\ref{fig:singlejet} and \ref{fig:gammajet} are summarized in Table~\ref{table}. 

As a comparison, results from LBT model simulations~\cite{Luo:2018pto,He:2018xjv} are also plotted in Figs.~\ref{fig:singlejet} and \ref{fig:gammajet} (red lines) and parameters $[\alpha,\beta,\gamma]$ from fits to the corresponding energy loss distributions are given in parentheses in Table~\ref{table}. 
LBT calculations can describe well the medium modification of single inclusive and $\gamma$-triggered jet spectra. Moreover, the jet energy loss distributions  $W_{AA}(x)$  from LBT which have a scaling behavior in $x=\Delta p_T/\langle \Delta p_T\rangle$ also agree well with the Bayesian extraction.  The effective mean number of out-of-cone scatterings is also small, which is consistent with the small extracted value of $\alpha$ as shown in Table~\ref{table}.  LBT results on the averaged jet energy loss at $\sqrt{s}=5.02$ TeV is slightly higher than Bayesian fits to the data, indicating the possible running of the effective strong coupling constant with the colliding energy.

Note that uncertainties as given by one-$\sigma$ variances for the parameters from fits to experimental data are quite large due to large experimental data errors and few data points, with the exception for $\gamma$-jet spectra at $\sqrt{s}=5.02$ TeV that have smaller errors and more data points. The large uncertainties come partially from the correlations of the extracted parameters as shown in Fig.~\ref{fig:parameter} for single jet analysis.  These uncertainties can be reduced as the quality of experimental data improves as one can see in the case of the Bayesian analysis of the LBT results. We should also note that CMS data on $\gamma$-jet spectra \cite{Chatrchyan:2012gt,CMS:2013oua,Sirunyan:2017qhf} have not been corrected for detector resolution as done in ATLAS experiment \cite{Aaboud:2018anc}, which can lead further corrections to the extracted jet energy loss distributions.

\begin{table}[h]
\begin{tabular}{|l|l|l|l|}
\hline
 \multicolumn{4}{|c|}{\centering single inclusive jet in $Pb+Pb$} \\
\hline
  & (0-10\%)2.76 TeV & (20-30\%)2.76 TeV &(0-10\%)5.02 TeV \\
\hline
$\alpha$ & \;$3.87\pm 2.93$  & \;$4.47 \pm 2.83$ & \;$4.41 \pm 2.86$ \\
&$(1.45\pm0.01)$ & $(1.33\pm0.02)$ & $(1.58\pm0.02)$\\
\hline
$\beta$ & \;$1.40 \pm 1.12$ & \;$1.12 \pm 0.47$ & \;$1.06 \pm 0.97$ \\
& $(1.39\pm0.06)$ & $(1.08\pm0.07)$ & $(1.56\pm0.06)$ \\
\hline
$\gamma$ & \;$0.21 \pm 0.09$  & \;$0.15 \pm 0.07$ & \;$0.26 \pm 0.06$ \\
& $(0.21\pm0.01)$ & $(0.20\pm0.01)$ & $(0.23\pm0.01)$ \\
\hline
 \multicolumn{4}{|c|}{\centering $\gamma$-triggered jet in $Pb+Pb$} \\
\hline
& (0-30\%)2.76 TeV & (30-100\%)2.76 TeV &(0-30\%)5.02 TeV \\
  \hline
$\alpha$ & \;$2.13 \pm 1.28$ &\; $3.75 \pm 2.81$ & \;$0.90 \pm 0.09$\\
& $(1.95\pm0.12)$ & $(1.04\pm0.06)$ & $(1.84\pm0.13)$\\
\hline
$\beta$ & \;$2.68 \pm 1.40$ & \;$0.55 \pm 0.44$ & \;$1.50 \pm 0.85$ \\
&$(0.72\pm0.06)$ & $(0.53\pm0.04)$ & $(0.50\pm0.04)$\\
\hline
$\gamma$ & \;$0.16 \pm 0.14$ & \;$0.13 \pm 0.18$ & \;$0.21 \pm 0.12$ \\
& $(0.44\pm0.02)$ & $(0.30\pm0.02)$ & $(0.56\pm0.02)$ \\
\hline
\end{tabular}
\caption{\label{table}
Parameters $[\alpha,\beta,\gamma]$ of the jet energy loss distribution from Bayesian fits to single inclusive and  $\gamma$-triggered jet spectra in $Pb+Pb$ collisions at $\sqrt{s}=2.76$ and 5.02 TeV.  Numbers in parentheses are from fits to LBT results.}
\end{table}

\noindent{\it Summary and Discussions}.--We have presented the first attempt to extract jet energy loss distributions from MCMC Bayesian analyses of experimental data on medium modification of both single inclusive and  $\gamma$-triggered jet spectra in $Pb+Pb$ collisions at $\sqrt{s}=2.76$ and 5.02 TeV. The energy loss distributions from both hard processes have a large width and the average jet energy loss increases with vacuum jet energy slightly faster than a logarithmic form. Results from LBT model simulations are consistent with the data-driven extraction and indicate that a small number of out-of-cone scatterings are responsible for the observed jet quenching. Reducing experimental uncertainties in finer $p_T$ bins should improve the precision of the Bayesian extraction. One can also consider flavor (quarks and gluons) dependence of the jet energy loss using pQCD calculations of the initial fraction of jet flavors.  Such systematic extraction of jet energy loss distributions can help to constrain model uncertainties in the study of jet transport coefficient and other properties of jet-medium interaction in high-energy heavy-ion collisions.

We thank T. Luo for providing LBT $\gamma$-jet results. This work is supported by DOE under Contract No. DE-AC02-05CH11231,  by NSF under Grant No. ACI-1550228 within the JETSCAPE Collaboration, by NSFC under Grants No. 11890714 and No. 11861131009. Computations are performed at GPU workstations at CCNU and DOE NERSC.



\end{document}